\documentclass{elsart}
\usepackage{graphicx,amssymb}
\usepackage[%
  breaklinks=true,%
  colorlinks=true,%
  linkcolor=blue,anchorcolor=blue,%
  citecolor=blue,filecolor=blue,%
  menucolor=blue,pagecolor=blue,%
  urlcolor=blue]{hyperref}

\newcommand{\ignore}[1]{}

\makeatletter
\def\elsartstyle{%
    \def\normalsize{\@setfontsize\normalsize\@xiipt{14.5}}
    \def\small{\@setfontsize\small\@xipt{13.6}}
    \let\footnotesize=\small
    \def\large{\@setfontsize\large\@xivpt{18}}
    \def\Large{\@setfontsize\Large\@xviipt{22}}
    \skip\@mpfootins = 18\p@ \@plus 2\p@
    \normalsize
}
\@ifundefined{square}{}{}
\makeatother

\begin{document}

\begin{frontmatter}
\title{Experimental demonstration of self-collimation beaming and splitting in photonic crystals}

\author{Aaron F. Matthews}
\ead{afm124@rsphysse.anu.edu.au}
\ead[url]{http://www.rsphysse.anu.edu.au/nonlinear}
\address{Nonlinear Physics Center and Center for Ultra-high bandwidth Devices for Optical Systems
(CUDOS), Research School of Physical Sciences and Engineering, Australian National
University, Canberra ACT 0200 Australia}

\begin{abstract}
I studied experimentally the beam self-collimation and splitting in two-dimensional microwave photonic crystals.  Using a microwave photonic crystal fabricated from alumina rods, I present an experimental proof of principle for an earlier theoretical proposal [A. Matthews et al., Opt. Commun. {\bf 279}, 313 (2007)] of a photonic crystal beam splitter based on the self-collimation effect.
\end{abstract}

\end{frontmatter}

\section{Introduction}
\label{intro}

Photonic crystals (PC) have been suggested and demonstrated in many studies
to make use of their photonic bandgap~\cite{Sakoda:2005:book}, such as defect structures, resonators, waveguides, couplers, splitters and other optical circuits~[2-4]. Another operating regime for PCs and related devices
is the anisotropic, in-band frequency domain, where substantial diffraction
control can be observed for propagating Bloch modes. In particular, the crystal
anisotropy leads to the well-known phenomenon of
{\em self-collimation}~[5-7],
that can be used for device applications such
as flat sub-wavelength lenses and wavelength-division
demultiplexers~[8-10].

Recently I, Matthews {\em et al.}~\cite{Matthews:2007-313:OC}, suggested a novel beam splitting
device based on the effect of self-collimation that provides large angular separation
using a small spatial footprint, and they highlighted the applicability of
this novel design of beam splitter to highly integrated photonic
systems. The main purpose of this paper is to study the {\em self-collimation effect
experimentally}, employing a two-dimensional microwave PC fabricated from alumina rods,
and to present an experimental proof of principle for the theoretical proposal of a PC beam splitter based on the self-collimation effect~\cite{Matthews:2007-313:OC}.


\begin{figure}[htb]
\centering\includegraphics[width=14cm]{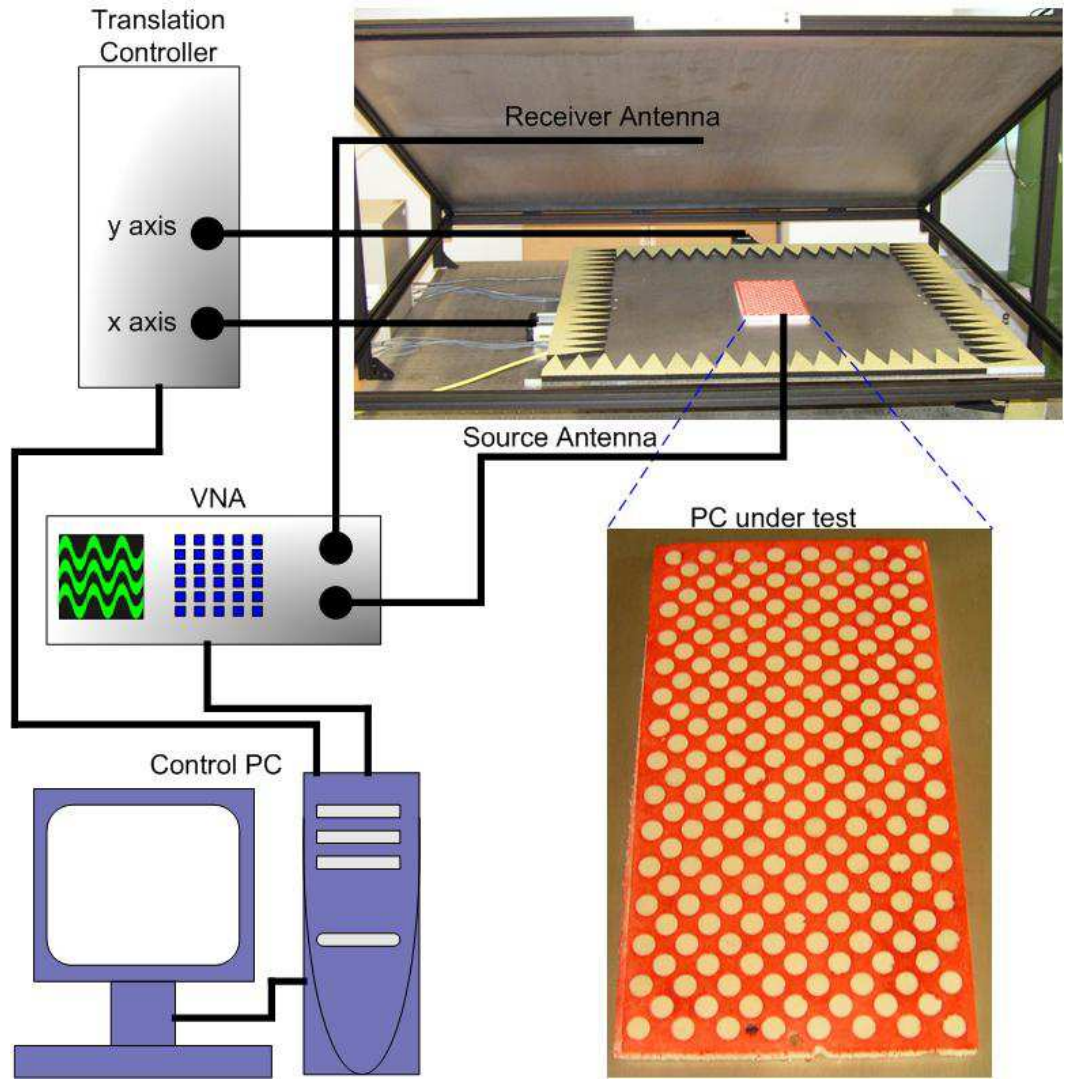}
\caption{(Color online) Schematic of the experimental setup used to map the field of the PC under test.
(Inset) A 2d PC formed from alumina rods mechanically supported by a frame of milled polystyrene foam.}
\end{figure}

\section{Experimental Design and Setup}

The design for the PC used in these experiments was produced using a combination of the planewave expansion method~\cite{Johnson:2000-173:OE}, implemented as Bandsolve, and the finite difference time domain (FDTD) method~\cite{Yee:1966-302:ITAP}, implemented as both Optiwave and Fullwave. The planewave expansion method was used to produce the isofrequency contours for the infinite PC structure which allow the determination of the frequency range of self-collimation for the base structure. The FDTD method was then used to model the specific structures to be prepared for the experiment. This confirmed that the fabricated structures would demonstrate my previous predictions~\cite{Matthews:2007-313:OC} within the experimental setup.

From these numerical designs a PC structure is fabricated. The structures used in this experiment were formed from high purity alumina ceramic rods, with an index of n$\approx$2.39 in the 2GHz to 11GHz frequency range, the operating range of the test setup. The rods used have a diameter of r=7.99mm$\pm$0.02mm and the PC has a lattice spacing of a=10.59mm. The rods are held by a frame of expanded polystyrene foam, n$\approx$1.01, manufactured for precision, using a computer controlled 3D Milling Machine (Roland DG MDX-40) into which the high purity alumina ceramic rods are mechanically inserted (Fig.1,inset). This gives a PC with a self-collimation frequency of 7.64GHz, equating to a wavelength of $\lambda$=39.24mm. In the case shown in figure 4 where an input waveguide is required, I used a machined block of polymethylmethacrylate (PMMA) (n$\approx$2.23), 1.27cm wide by 13cm long by 1cm high.

Figure 1 shows the experimental setup. The manufactured PC is placed in a parallel plate waveguide with a separation between the plates of $\sim$11mm. A source antenna of 1.3mm diameter and 10mm height is set in the lower plate, and a detection antenna, consisting of a coaxial cable, is set flush in the upper plate. The structure operates in a fundamental TEM mode over the 2GHz to 11GHz frequency range, which is well suited to modelling as a 2D TE PC system. Shaped carbon impregnated foam surrounds the measurement region to suppress reflections from the boundaries of the waveguide. The upper plate is held firmly in place, with the detection antenna static, and the lower plate, including both the input antenna and the PC, mounted on a planar translation stage (Parker, 803-0936A, two-axis translation stage) to allow scanning of the field from the top surface of the PCs. A vector network analyzer (VNA), Rohde and Schwarz model ZVB20, is used to provide both the signal to the input point source and receive the output from the receiving antenna. The magnitude and phase of the transmission response, S$_{12}$, are measured using the VNA, and the orientation of the antenna means that this response is highly correlated to the vertical component of the electric field (equivalent to the TE mode).

\section{Self-collimation in photonic crystals}

\begin{figure}[htb]
\centering\includegraphics[width=11cm]{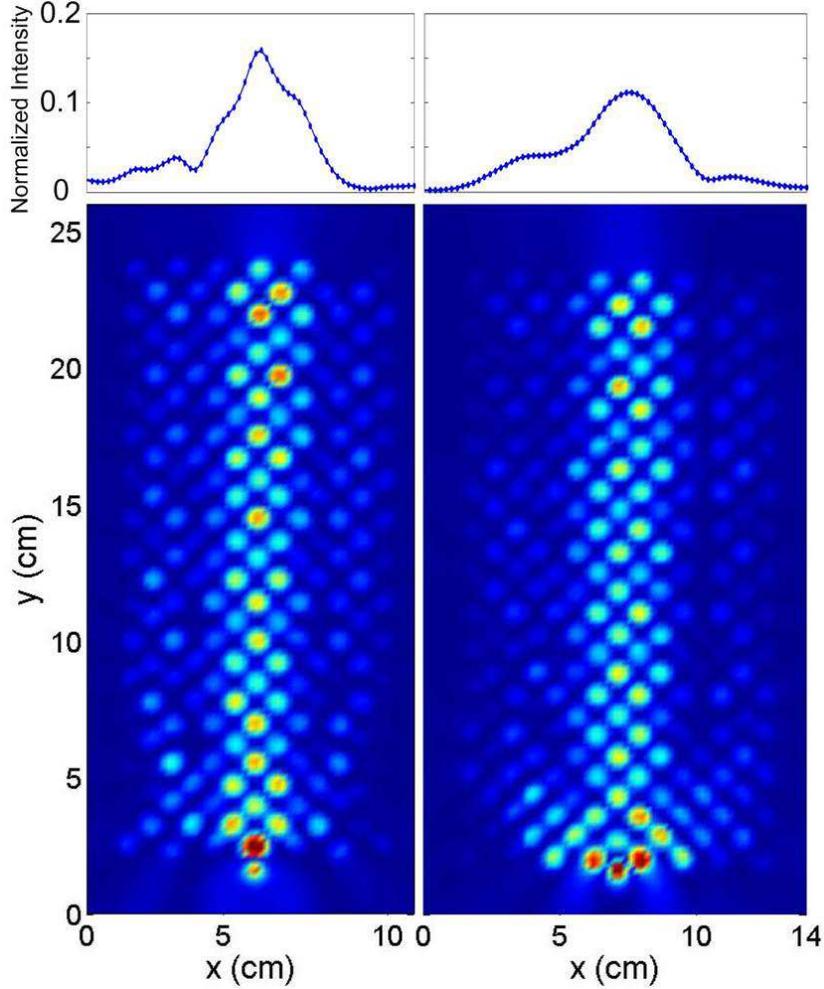}
\caption{(Color online) Experimental long range self-collimation stimulated from a point source (Left) source on-site, adjacent to the center surface rod, and (Right) off-site, between the two central surface rods. (Top) Slices of the field at the output from the PCs taken one lattice constant(a) from the surface. Output slices are normalized to the input point source. (Bottom) The scanned field measured above the experimental PC.}
\end{figure}

To demonstrate self-collimation I orient a square lattice PC of alumina rods along the $\Gamma$-M direction, the self-collimation direction for the lowest order mode. I form a rectangle of 13 by 29 rods (Fig. 2), in the first case, for the on-site stimulation, where the input point source is adjacent to the central base rod. In the second case, off-site stimulation where the structure is 15 by 29 rods with the source between the central two base rods. The difference in base widths is to allow symmetry around the source.

Figure 2 shows in both cases strong self-collimation observed at 7.64GHz as predicted. The field in the on-site case is seen to be confined within the central 3 rows of rods confirming visually a reduction of the diffraction. In the off-site case, the initial coupling into the perpendicular self-collimation directions leads to an increase in the post-coupled beam width increasing it to 4 rods wide. This results in a broadened output profile compared to the on-site case. The asymmetry and lateral shift in the field peak, seen in the output, in both cases, is a result of the initial misalignment of the structure. The beams can be seen to retain a full width half maximum in both cases of under 3cm after 22cm of propagation.

This result is particularly noteworthy as, I understand, it is the longest distance over which the self-collimation of electromagnetic radiation has been observed. It is also worth noting that I am able to visualize the field in-situ showing in the clearest manner possible the removal of diffraction in a self-collimating PC.
With self-collimation in this structure confirmed I am able to move on and consider the rotated PC structure where an input beam is coupled to two separate self-collimation directions in the form of a beam splitter.

\section{Self-collimated beaming splitter}
\begin{figure}[htb]
\centering\includegraphics[width=14cm]{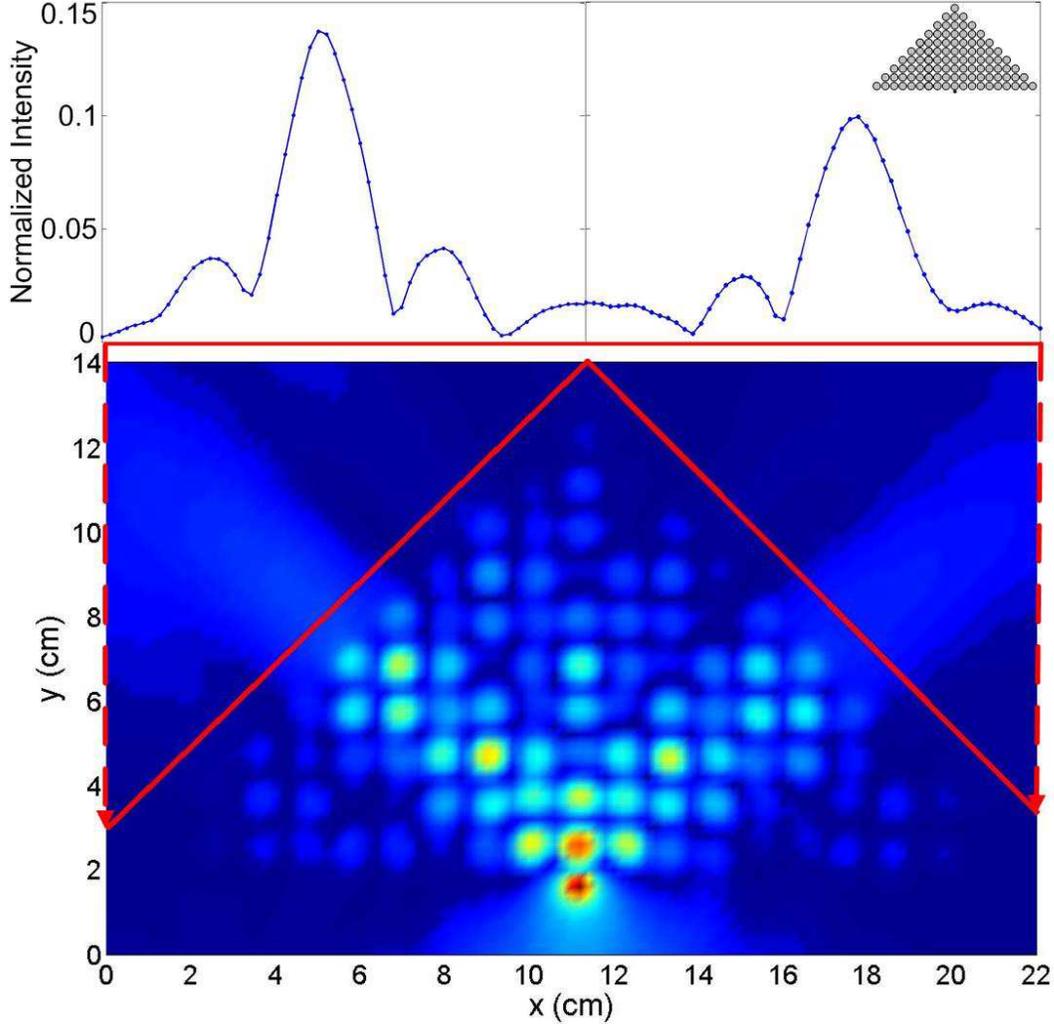}\label{fig3}
\caption{(Color online) Experimental observation of self-collimation beam splitting. Inset: A triangular square lattice PC is stimulated by a point source on-site with the central surface rod. The outputs line plots (top) are measured along the red lines in the field plot (bottom).}
\end{figure}

\begin{figure}[htb]
\centering\includegraphics[width=14cm]{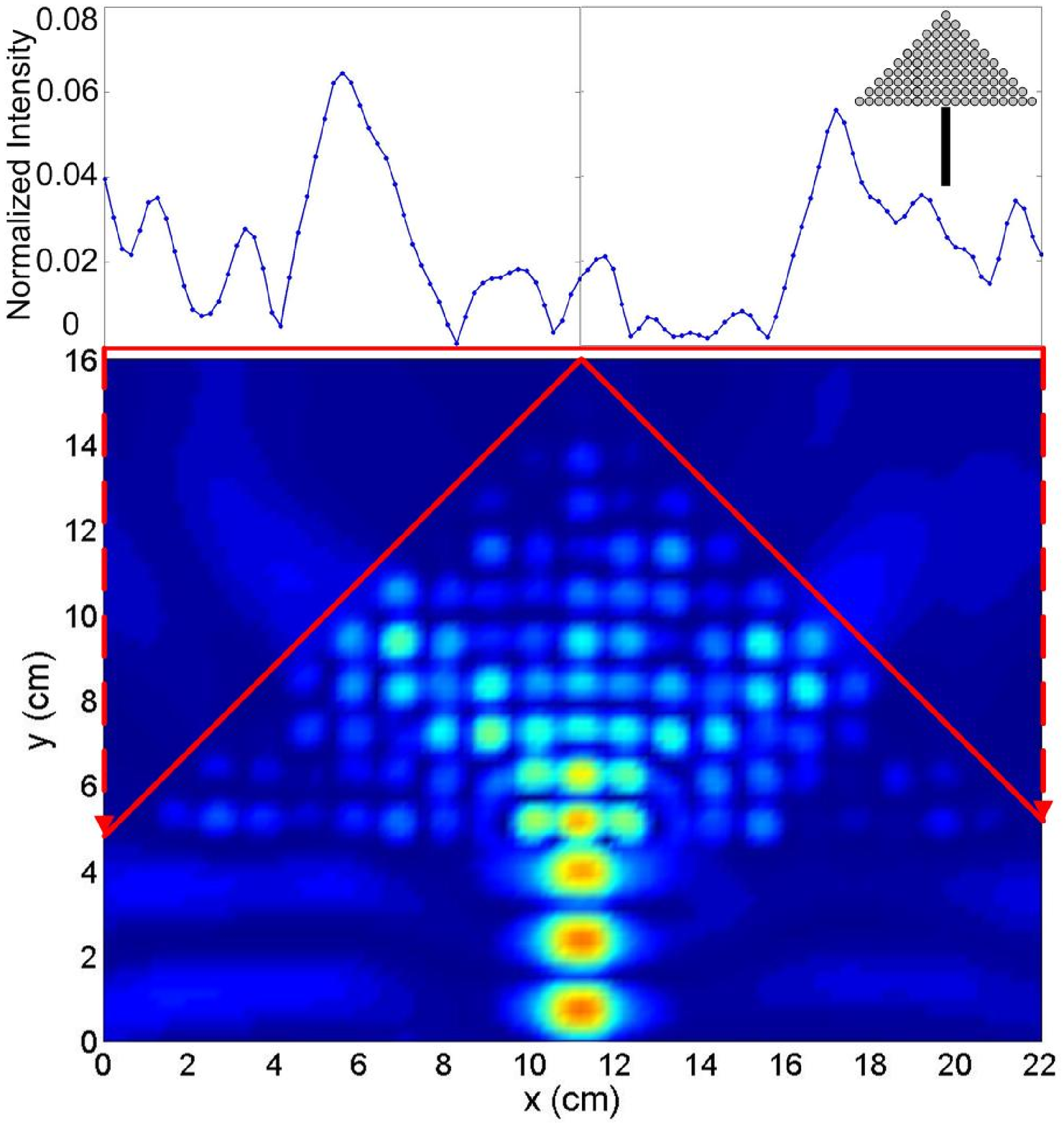}
\caption{(Color online) The same structure as in figure~\ref{fig3} except in this case stimulated from an input waveguide (1.27cm wide by 13cm long by 1cm high), rather than a point source.}
\end{figure}

By rotating the PC from the $\Gamma$-M direction (the self-collimation direction) to the $\Gamma$-X direction I am able to stimulate two collimation directions for a single input with a central directional bandgap. Using this property I am able to obtain a symmetrical beam splitter design. This design provides a full 90-degree beam separation with an implicit symmetry of the two outputs. The other advantage of such a design is the small spatial footprint which will be a particular advantage when the device is scaled to integrate in a photonic chip.

I implement such a beam splitter in the aforementioned alumina PC lattice. It is formed as a triangle of rods oriented in the $\Gamma$-X direction with a base width of 19 rods and depth of 10 rods (Fig. 3,inset). The PC is stimulated at the self-collimation frequency (7.64GHz) from a point source adjacent to the central base surface rod  and observe splitting into two of the self collimation, $\Gamma$-M, directions of the PCs and a directional bandgap in the forward $\Gamma$-X direction.

In figure 3, we observe that with the full range of wave vectors stimulated beam splitting is indeed obtained. From this we obtain the expected output profile (Fig. 3) with a pair of distinct peaks (the peak field is normalized to the input point source field) separated by a central null. The self-collimation is observed in the confinement of the two beams in 4 rod widths and the beams retain their form after coupling out of the PC.

This result confirms the predictions of our earlier theoretical work~\cite{Matthews:2007-313:OC} and shows that such a system could be scaled to work at optical frequencies.
The idea of coupling from a point source becomes unreasonable because it would have to be integrated in a photonic chip, therefore coupling from a waveguide should be considered.

In figure 4, the point source has been replaced with a PMMA waveguide 1.27cm wide by 13cm long by 1cm high. While the output from a waveguide lacks the full range of wave vectors it can be seen that the interface with the PC facilitates the coupling of the power into the self-collimation directions.
The result of this is the expected two peak output with the wide null in the forward direction.

In both beam splitting cases, disorder appears from the experimental setup. In particular, the minor ellipticity in the rods, the tapering of the width along the height of the rods, and the structural uncertainty in the position of the rods leads to significant asymmetry in the output beam. The addition of a waveguide in figure 4 exacerbates this problem, increasing the sensitivity of the structure to these disorders.

With this result I have confirmed that such a beam splitter coupled to a waveguide is indeed able to be incorporated in a photonic chip. I also note that this is the first time that such a beam splitter integrated with a waveguide has been demonstrated.

\section{Conclusions}

I have studied experimentally the beam self-collimation and splitting of self-collimated beams
in two-dimensional microwave PCs and confirmed our theoretical proposal for
a PC beam splitter based on the self-collimation effect~\cite{Matthews:2007-313:OC}.
I investigated the specific properties of the self-collimation effect using microwave
PCs fabricated from alumina rods, and demonstrated that the self-collimation based beam
splitter provides a large angular separation between the output beams using a small spatial footprint.
I believe that this kind of beam splitter, even being realized for microwaves, may provide significant advantages for optical integration and routing based on self-collimation and this concept can be demonstrated in the near future for the optical domain.

The author thanks S.K. Morrison, D.A. Powell, I.V. Shadrivov, and Yu.S. Kivshar for useful discussions and suggestions. This work has been supported by an award under the Merit Allocation Scheme on the National Facility of the Australian Partnership for Advanced Computing and by the Australian Research Council through the Center of Excellence Program.

\end{document}